\documentstyle[11pt,newpasp,twoside,epsf]{article}
\newcommand{\civ}{C\,{\sc iv}}

\def\edcomment#1{\iffalse\marginpar{\raggedright\sl#1\/}\else\relax\fi}
\reversemarginpar

\begin{document}
\title{Broad absorption line quasars have the same cool dust emission as quasars without BALs}
\author{Chris J. Willott}
\affil{Herzberg Institute of Astrophysics, National Research Council,\\
5071 West Saanich Rd, Victoria, B.C. V9E 2E7, Canada}
\author{Steve Rawlings, Jennifer A. Grimes}
\affil{Astrophysics, Department of Physics, Keble Road, Oxford, OX1
3RH, U.K.}

\begin{abstract}

The results of a sub-millimeter survey of SDSS broad \civ\ absorption
line quasars is discussed. It is found that the sub-millimeter flux
distribution of BAL quasars is similar to that of non-BAL
quasars. This is consistent with the idea that all quasars contain
broad absorption line regions, but only a fraction of them are visible
along our line-of-sight. The observations are inconsistent with BAL
quasars being observed at a special evolutionary epoch co-inciding
with a high star-formation rate and dust mass.
\end{abstract}

\section{Introduction}
About 15\% of optically-selected quasars show metal absorption systems
with huge blue-shifted velocities of several thousand kms$^{-1}$.
These broad absorption lines (BALs) are attributed to outflows located
close to the accretion disk. The similarity of the continuum and line
emission of BAL and non-BAL quasars motivates the hypothesis that BAL
quasars are not intrinsically different from other quasars. The
presence of BAL features in a fraction of quasars can be explained by
a difference in viewing angle if the sub-relativistic outflow at the
origin of the BAL features is not isotropic (Weymann et al. 1991). The
popular notion that the outflow is preferentially located close to the
edge of the torus surrounding the supermassive black hole has found
support in spectropolarimetric measurements (Goodrich \& Miller 1995;
Cohen et al. 1995). However, studies of radio-loud BAL quasars, in
which the radio properties give some information on the orientation,
appear at odds with this simple orientation model (Becker et
al. 2000). The other main contender as an interpretation of the BAL
phenomenon is the evolutionary scenario of Briggs et al. (1984) in
which all quasars pass through a BAL phase for $\sim$ 15\% of their
active lifetimes. The small sizes of the radio lobes in radio-loud BAL
quasars (Becker et al. 2000) is suggestive of the BAL phase
co-inciding with an early stage of quasar activity -- perhaps removing
a shroud of gas and dust from the nuclear region.

The sub-millimeter emission from quasars comes from optically-thin, cool dust
and is therefore orientation-independent. It is also likely heated by
young stars in starbursts which will evolve over the lifetime of the
quasar. It is therefore the prime wavelength in which to discriminate
between the orientation and evolutionary explanations of the BAL
phenomenon. If all quasars contain BALs, then BAL quasars would be
expected to have the same sub-millimeter properties as non-BAL quasars. But if
the BAL phenomenon is a phase that all quasars go through, and is
connected with the termination of large-scale star-formation, then the
sub-millimeter emission should differ between BAL and non-BAL quasars.
It is assumed that $H_0\,=\,70\, {\rm km\,s^{-1}\,Mpc^{-1}}$,
$\Omega_{\mathrm M}=0.3$, $\Omega_\Lambda=0.7$. 

\section{A sub-millimeter survey of broad absorption line quasars}

We have observed with the JCMT a sample of BAL quasars selected
from the SDSS EDR BAL sample of Reichard et al. (2003). The
sample was designed to be as similar as possible to BAL quasars which
do not show broad absorption lines which had already been observed at
either the JCMT (Priddey et al. 2003) or IRAM (Omont et
al. 2003). Here we briefly summarize the BAL sample selection:
redshift range $2<z<2.6$, RA range 10$^{\rm h}$ - 18$^{\rm h}$, \civ\
balnicity index BI (Weymann et al. 1991) $>200\,{\rm km\, s}^{-1}$,
dereddened absolute $B$-band $M_B<-26.6$. Full details
of this project are given in Willott, Rawlings \& Grimes (2003). 

The entire sample of 30 BAL quasars was observed in photometry mode
with the SCUBA bolometer array at the JCMT from February to June 2003.
We achieved our aim of reaching an rms sensitivity at 850\,$\mu$m
lower than 3.33 mJy (i.e. $3\sigma=10$\,mJy) for all the BAL quasars
observed. This sensitivity matches that obtained in the observations
of the non-BAL comparison sample. Eight BAL quasars were detected at
the $>2 \sigma$ level (four of which are securely detected at the $>3
\sigma$ level).  The detected quasars span a range of $z$, $M_B$ and
balnicity index. Although one can use the observed sub-mm flux to
obtain an estimate of the far-IR luminosity, there are many
uncertainties in this calculation. Since the relationship between flux
and luminosity does not change much over the redshift range we are
interested in for realistic dust spectral energy distributions, we
will use the flux rather than luminosity in further analysis.

\begin{figure}
\plotfiddle{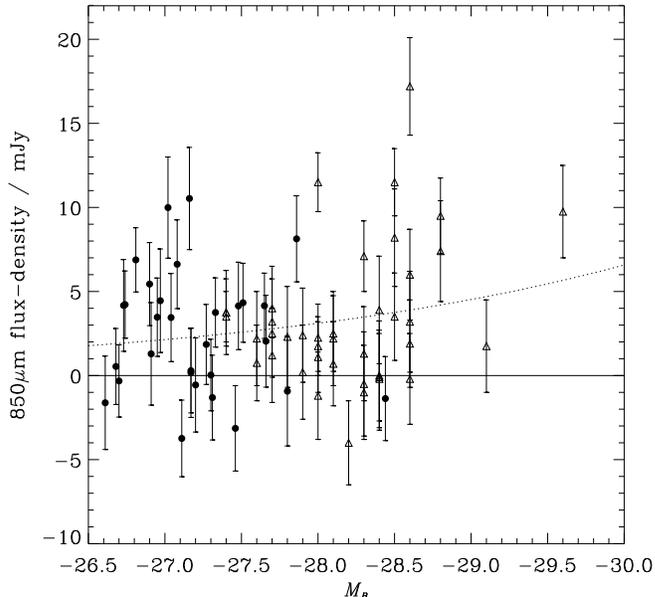}{0.0cm}{0}{42}{42}{-135}{-216}
\vspace{7.1cm}
\caption{850\,$\mu$m flux-density against absolute magnitude for the
BAL quasar sample (filled circles) and the non-BAL control sample
(open triangles). Error bars show the $1 \sigma$ uncertainty on the
measured fluxes. The dotted line illustrates a positive correlation between
sub-millimeter flux and optical luminosity where a difference in $M_B$
of 1 equals a change in $S_{850}$ of a factor of $1.5$.}
\end{figure}

\section{Comparing the sub-millimeter fluxes of BAL and non-BAL quasars}

We carried out a variety of statistical tests on the sub-mm flux
distributions of the BAL and non-BAL quasar samples. The aim was to
determine whether there is any difference between these two types of
quasar. The tests included sample statistics, such as the median, mean
and weighted mean flux of each sample. No evidence for a difference
between BAL and non-BAL quasars was found in these tests. The mean
850\,$\mu$m fluxes of the BAL quasars is $2.56 \pm 0.67$ mJy and for
the non-BAL quasars it is $3.34 \pm 0.64$ mJy.

Due to the large number of 850\,$\mu$m upper limits in both samples,
we also use survival analysis statistical tests which can account for
these limits (Feigelson \& Nelson 1985; Isobe, Feigelson \& Nelson
1986). To minimise the number of upper limits, we consider sources
with snr $>2$ to be detections for these tests. We wish to determine
whether the 850\,$\mu$m flux distributions of the BAL and non-BAL
quasars are different. Using a variety of tests (the Gehan, logrank
and Peto-Prentice tests), the returned probabilities that the fluxes
of the BAL and non-BAL samples are drawn from the same distribution
lie in the range (54-99\%). Again, we find no evidence that BAL and
non-BAL quasars differ in their sub-millimeter fluxes.

In Fig. 1 we plot the sub-millimeter fluxes of the BAL and non-BAL
quasars as a function of their absolute optical magnitude. Due to the
fact that the non-BAL quasars come from samples with a brighter
optical magnitude limit than the SDSS, the absolute magnitudes of the
non-BAL quasars are brighter than those of the BAL quasars by about
one magnitude. We have to be careful that a correlation between
sub-millimeter flux (luminosity) and optical luminosity is not
affecting our results. For example a positive correlation between
optical and sub-millimeter luminosity combined with enhanced
sub-millimeter luminosity in BAL quasars could have led to the
observed result that the two samples have indistinguishable
sub-millimeter flux distributions. 

The question of whether there is a correlation between the
sub-millimeter and optical luminosities has been addressed before
(Priddey et al. 2003; Omont et al. 2003) and the consensus is that
there is no measurable correlation. We use the survival analysis
bivariate correlation tests to assess the signficance of any
correlations in these datasets. We find no significant correlation in
the BAL quasar sample, but a very significant anti-correlation in the
non-BAL sample of 41 quasars. The reason that we have now detected
this positive correlation between the sub-millimeter and optical
luminosities may be that by combining the samples over a restricted
redshift range we have disentangled luminosity and redshift effects.

For the purposes of our BAL quasar survey we wish to know if the
existence of a correlation between $S_{850}$ and $M_B$ could affect
our findings. The median absolute magnitude of the BAL sample is
$M_B=-27.2$ and for the non-BAL sample it is $M_B=-28.2$. Therefore,
we scale the observed sub-millimeter fluxes of the BAL sample by a
factor of 1.5 (except for those with negative flux) to simulate the
effects of a correlation with slope similar to that observed in the
non-BAL sample (and plotted in Fig. 1). We repeated the two sample
tests with these scaled BAL fluxes and the comparison sample of
non-BAL quasars. The probabilities that the sub-millimeter
fluxes of the BAL and non-BAL samples are drawn from the same
distribution are 14\% and 27\%.  This test shows that, even after
allowing for the effects of a weak correlation between sub-mm flux and
$M_B$, there is not a significant difference between BAL and non-BAL
quasars.

Our observations and analysis have shown that quasars with broad
absorption lines show the same sub-millimeter properties as quasars
without broad absorption lines. What does this tell us about the
nature of the BALs and the evolution of quasars? Since sub-millimeter
emission is optically-thin there should be no orientation dependence
of sub-millimeter emission. Our finding that the sub-millimeter
emission is not related to the presence of a BAL is therefore
consistent with the orientation hypothesis that all quasars contains
BALs, but only a fraction of them are visible along our line-of-sight.
Our results are inconsistent with a model in which the presence of
BALs is associated with a special epoch in quasar/galaxy evolution
co-inciding with a large dust mass and a high star-formation
rate. However there are other evolutionary models which could be made
consistent with our data. For a fuller description of this project and
its implications the interested reader is referred to Willott et
al. (2003).

\end{document}